\documentclass[12pt,a4paper]{article}
\usepackage{graphics}
\usepackage{a4}
\usepackage{amstex}
\usepackage{times}
\title{Noisy regression and classification 
with continuous multilayer networks}
\author{M. Ahr$^{1}$, M. Biehl$^{1}$, and R. Urbanczik$^{2}$ \\ \\
$^{1}$Institut f\"ur Theoretische Physik \\ 
Julius-Maximilians-Universit\"at W\"urzburg \\
Am Hubland, D - 97074 W\"urzburg, Germany \\
$^{2}$Neural Computing Research Group \\
Aston University \\
Aston Triangle, Birmingham B4 7ET, UK}

\begin{document}
\newcommand{\sign}{\mbox{sign}}
\newcommand{\g}{\mbox{g}}
\setlength{\unitlength}{0.1\textwidth}
\maketitle
\begin{abstract}
We investigate zero temperature Gibbs learning for two classes of unrealizable 
rules which play an important r\^{o}le in practical applications of 
multilayer neural networks with differentiable activation functions: 
classification problems and noisy regression problems.  
Considering one step of replica symmetry breaking, we surprisingly
find that for sufficiently large training sets the stable state
is replica symmetric even though the target rule is unrealizable.
Further the classification problem is shown to be formally equivalent
to the noisy regression problem.
\end{abstract}

Neural networks with differentiable activation functions play an important 
r\^{o}le in practical applications \cite{Hertz:Krogh:Palmer}.
Besides being used for regression, they are
often applied to classification problems as well, since 
gradient based methods are available for training such networks. 
In both cases, given a training set
of $P$ input/output pairs $(\vec{\xi}_{\mu},\theta_{\mu})$, 
$\vec{\xi}_{\mu} \in \mbox{I\!R}^{N}, \theta_{\mu} \in \mbox{I\!R}$, one 
adapts the network with output $\sigma$ to minimize a cost function which
measures the deviation between 
$\sigma(\vec{\xi}_{\mu})$ and the target output $\theta_{\mu}$.

For the {\em regression problem} we shall assume that the target
output is a function $\tau$  of the input, corrupted by additive noise,
so $\theta_{\mu} := \tau(\vec{\xi}_{\mu}) + \gamma \nu_{\mu}$.
The noise terms $\nu_{\mu}$ are independent and normally distributed.
An appropriate cost function then is the quadratic error 
 \begin{equation}
H = P \epsilon_{t} = 
\frac{1}{2}\sum_{\mu = 1}^{P} \left( \sigma(\vec{\xi}_{\mu}) - 
\theta_{\mu} 
\right)^{2} \, . 
\label{energie}
\end{equation}

We call $\epsilon_{t}$, the mean energy per example, {\em training error}.
The main goal of learning, however, is to minimize 
the {\em prediction error} $\epsilon_{p}$, defined as the expectation 
value of the training error on a new example, that is $\epsilon_{p} = 
\left<(\sigma(\vec{\xi}) - \theta(\vec{\xi}))^{2}\right> / 2$, where the 
average is performed 
over the distribution of inputs and the randomness of $\theta$ in the 
presence of noise.

In {\em classification problems} only a binary label is available for
the examples and we shall assume that $\theta_{\mu} = 
\lambda\, \sign(\tau(\vec{\xi}_{\mu}))$. Here $\tau$ is some function of the
input and $\lambda$ is a tunable parameter. 
One is then mainly  interested in the sign of the networks output,
that is the goal of learning is to minimize the {\em classification error}
\begin{equation}
\epsilon_{c} = \left< \Theta( - 
\sigma(\vec{\xi}) \tau(\vec{\xi}) ) 
\right> \; ,
\end{equation}
where $\Theta$ is the Heavyside step function.
However the empirical mean of this performance measure
\begin{equation}
P^{-1} \sum_{\mu = 1}^{P} \Theta( - \sigma(\vec{\xi}_{\mu})\theta_{\mu})
\label{emp}
\end{equation}
is piecewise constant and cannot be optimized using e.g. backpropagation.
While the sample complexity of training multilayer networks based on 
(\ref{emp}) has been analysed in \cite{Schwartze,Schwartze:Hertz,Urbanczik},
practical applications of neural networks 
\cite{Hertz:Krogh:Palmer,Bishop,Eisenstein:Kanter}
typically use the differentiable cost function (\ref{energie}) even for 
classification tasks. So for the purposes of training, classification is
mapped onto regression, and the question arises how this affects the 
generalization behaviour.   
(Alternative cost functions have been studied 
in the context of online learning \cite{Barkai:Seung:Sompolinsky}.)

Here we present a theoretical investigation of the two learning problems. 
We focus on a simple two-layered student network
which consists of $K$ hidden units with activation 
function $\g(x) = \mbox{erf}(x/\sqrt{2})$ and $N$-dimensional 
weight vectors $\{\vec{J}_{i}\}_{i = 1}^{K}$, where $\vec{J}_{i}^{2} = N$. 
The output unit is linear and has weights fixed to the
value $1/\sqrt{K}$. Then, the output of this network which is called
``soft-committee machine'' \cite{Saad:Solla,Kang:Oh:Kwon:Park} is
\begin{equation}
\sigma(\vec{\xi})
 = \frac{1}{\sqrt{K}} \sum_{i = 1}^{K} \g\left(\frac{\vec{J}_{i} \cdot 
\vec{\xi}}{\sqrt{N}} \right) \ .
\end{equation}

The target function $\tau(\vec{\xi})$ will be given by
a soft-committee machine with the same 
number of hidden units as the student network and weight vectors
$\{\vec{B}_{i}\}_{i = 1}^{K}$, where 
$\vec{B}_{i} \cdot \vec{B}_{j} = N \delta_{ij}$. 
So the classification problem is perfectly learnable 
in the sense that the student network can achieve $\epsilon_{c} = 0$
if its weight vectors become identical to those of the teacher network. 
Further, we assume the components of the examples to be independent
random numbers with mean zero and unit variance. 

We use the well-known replica formalism to investigate these
problems in the 
thermodynamic limit $N \rightarrow \infty$. This requires the 
calculation of the quenched free energy
\begin{equation} 
F = -\frac{1}{\beta}  \left< \ln Z \right> 
= - \frac{1}{\beta} \left. \frac{\partial}{\partial n}\ln \left< Z^{n}
\right> \right|_{n = 0}
\end{equation}
where $Z^{n}$ is the partition function
$\int d\mu(\{\vec{J}_{i}^{a}\}) 
\exp(- \beta \sum_{a = 1}^{n} H(\{\vec{J}_{i}^{a}\}_{i = 1}^{K}))$
of $n$ replicas (labeled $a, b = 1, 2, 3, \ldots$)
of the student network \cite{Seung:Sompolinsky:Tishby,Watkin:Rau:Biehl}. 
Here $H$ is 
interpreted as the energy of a system which is in thermal equilibrium 
at a temperature $T = 1/\beta$.
In the limit of zero temperature, $\beta \rightarrow \infty$, $F$ is the 
optimal value of the energy which can be achieved by minimizing
$H$ with respect to the network weights.

Introducing an additional integration over the {\em order parameters} 
$Q_{ij}^{ab} := \vec{J}_{i}^{a} \cdot \vec{J}_{j}^{b} /N$
and $R_{ij}^{a} := \vec{J}_{i}^{a} \cdot \vec{B}_{j}/N$, which is 
performed as a saddle point integration in the limit of large $N$, 
we find for moments of the partition function,
$\ln\langle Z^{n}\rangle = -N (\alpha K G_{r} + s)|_{\mbox{extr}}$. 
Here $G_{r}$ is an effective Hamiltonian and the entropy term 
$s = (1/2) \ln \det \underline{\underline{C}}$,
where 
$\underline{\underline{C}}$ is the $K(n + 1) \times K(n + 1)$-dimensional 
matrix of the order parameters \cite{Ahr:Biehl:Urbanczik}.
We have further introduced 
the rescaled number of examples, $\alpha = P/(N K)$.  

In the following we restrict ourselves to the limit of large $K$.
We make a site-symmetric Ansatz for the dependence of the order parameters
on the site indices $i, j$:
\begin{eqnarray}
R_{ij}^{a} & = & \delta_{ij} \left( \frac{\hat{R^{a}}}{K} + \Delta^{a} \right)
+ ( 1 - \delta_{ij}) \frac{\hat{R^{a}}}{K}  \\
Q_{ij}^{ab} & = & \delta_{ij} \left( \frac{\hat{Q}^{ab}}{K} + \delta^{ab} 
\right) +
(1 - \delta_{ij}) \frac{\hat{Q}^{ab}}{K}
\end{eqnarray} 
The scaling of the unspecialized order parameters with the number 
of hidden units 
results from the condition that the outputs of the students $\sigma^a$ must 
be of order $1$ in the limit of infinite $K$.
In this limit the calculation of 
$G_{r}$ can be carried out analytically since the joint
distribution of the  $\sigma^{a}$ and $\tau$ becomes Gaussian 
\cite{Urbanczik}.

For the regression problem, using a one step replica symmetry breaking Ansatz,
we obtain in the limit $n \rightarrow 0$:
\begin{equation} G_{r}^{0} = \frac{1}{2} \left( \frac{X_{1}}{X_{2}} +
\frac{m - 1}{m} \ln X_{3} + \frac{1}{m} \ln X_{2} \right) \; ;
\label{grnull}
\end{equation}
where 
\begin{eqnarray*}
X_{1} & = & \beta( v^{0} - 2 w + 1/3 + \gamma^{2})\; \\
X_{2} & = & 1 + \beta(u + (m - 1)v^{1} - m v^{0})\; \\
X_{3} & = & 1 + \beta(u - v^{1})\; .
\end{eqnarray*}
$u = 1/3 + \hat{Q}/\pi$, $v^{1} = f(\delta^{1}, \hat{Q}^{1})$, 
$v^{0} = f(\delta^{0}, \hat{Q}^{0})$, and $w = f(\Delta, \hat{R})$ are 
the covariances of the $\sigma^{a}$ and $\tau$, where 
$$ f(x, y) = \frac{2}{\pi} \arcsin \left(\frac{x}{2}\right) + \frac{y}{\pi} 
\; .$$
It is easy to calculate $s$ and to perform the limit $n \rightarrow 0$
to obtain the entropy term $s^{0}$ in the free energy which is the
same as for hard committee machines \cite{Urbanczik}.
The order parameter $\Delta$ indicates specialization of the network: 
if $\Delta = 0$, the network configuration
is unspecialized, i.e. a weight vector of the student network has the 
same overlap ($\hat{R}/K$) 
with all weight vectors of the teacher network, whereas 
a positive $\Delta$ indicates a specialized configuration where each 
of the student vectors has a greater overlap ($\hat{R}/K + \Delta$)
with one of the teacher 
vectors than with the others. 
$\hat{Q}/K$ is the cross-overlap between different weight vectors of a student.
The remaining order parameters $\hat{Q}^{0}$, $\hat{Q}^{1}$, 
$\delta^{0}$, $\delta^{1}$ and $m$ parametrize the distribution of overlaps 
between the weight vectors of different students.
Note that as in \cite{Ahr:Biehl:Urbanczik}, 
using the saddle point equations for the free energy, one 
may analytically eliminate the unspecialized order
parameters  $\hat{R}$, $\hat{Q}$, $\hat{Q}^{0}$ and $\hat{Q}^{1}$.

In terms of the order parameters the prediction error for the regression 
problem is given by:
\begin{equation}
\epsilon_{p} = \frac{1}{3} + \frac{\hat{Q}}{2 \pi} - \frac{\hat{R}}{\pi}
- \frac{2}{\pi} \arcsin \left(\frac{\Delta}{2}\right) + \frac{\gamma^{2}}{2}\;.
\end{equation}

The replica calculation for the classification problem is 
analogous. It yields the same entropy $s^{0}$ and 
a $G_{r}^{0}$ of the form (\ref{grnull}) with identical $X_{2}$ and $X_{3}$
but 
\begin{equation}
X_{1} = \beta (v^{0} - 2 w \lambda \sqrt{6/\pi} + \lambda^{2})\;.
\label{grnullc}
\end{equation}
For the prediction  and classification error one finds 
\begin{eqnarray}
\epsilon_{p} &=& \frac{1}{6} + \frac{\hat{Q}}{2 \pi} - 
\lambda \sqrt{\frac{6}{\pi}} \left( \frac{2}{\pi} \arcsin 
\left(\frac{\Delta}{2}\right)
+ \frac{\hat{R}}{\pi} \right) + \frac{\lambda^{2}}{2} 
\label{epsilonp} \\
\epsilon_{c} & = & \frac{1}{\pi} \arccos \left[
\left( \frac{2}{\pi} \arcsin \left(\frac{\Delta}{2}\right) + \frac{\hat{R}}{\pi} 
\right) \left( \frac{1}{9} + \frac{\hat{Q}}{3 \pi} \right)^{-\frac{1}{2}}
\right]\;.
\end{eqnarray}
In the limit of large sample size $P$ the training error   
$\epsilon_{t}$ will converge to $\epsilon_{p}$. 
So for the classification error to become zero the value of $\lambda$
must be chosen so that the minima of $\epsilon_{p}$ and $\epsilon_{c}$
coincide.
Note that the order parameters are constrained by the 
fact that the vectors $\vec{a} := (1/N) \sum_{i = 1}^{K} \vec{J}_{i}$ and 
$\vec{b} := \vec{B}_{j}$
must fulfill $(\vec{a} \cdot \vec{b})^{2} \leq \vec{a}^{2} \vec{b}^{2}$, 
which demands
$\hat{Q} \geq (\Delta + \hat{R})^{2} - 1$. Minimizing the prediction
error (\ref{epsilonp}) under this restriction,
we obtain $\Delta = 1$ and
$\hat{Q} = \hat{R} = 0$ (student and teacher network identical, 
$\epsilon_{c} = 0$) 
only for $\lambda = \lambda_{o} = \sqrt{\pi/6}$. This is the optimal value of 
$\lambda$ which allows asymptotically perfect  
classification. Inserting $\lambda = \lambda_{o}$ in 
equation (\ref{grnullc}) and comparing to (\ref{grnull}), 
one finds that in this case the free energy of the 
classification problem is identical to that of a noisy regression problem 
with $\gamma = \gamma_0 = \sqrt{\pi/6 - 1/3}$. In the sequel we shall
only consider the case $\lambda = \lambda_0$ for the classification problem.

We  focus on the limit of zero temperature and the construction of
this limit depends on whether a zero training error is achievable.
Denoting this critical capacity by $\alpha_{\rm c}(\gamma)$, we find
that $\alpha_{\rm c}(\gamma)$  decreases to $0$ with increasing 
$\gamma$ and $\alpha_{\rm c}(\gamma)\rightarrow 1$ as $\gamma\rightarrow 0$. 
This is explained by the fact that the noise increases the magnitude 
of the target outputs. This correlates the hidden units of the student
and thus reduces the storage capacity.

Below $\alpha_{\rm c}(\gamma)$ we find an unspecialized replica symmetric
solution with $\Delta=\delta^1=\delta^0=0$. Above $\alpha_{\rm c}(\gamma)$
one finds  $\delta^{1}\rightarrow 1$ for $\beta\rightarrow\infty$ and 
the appropriate scaling
is $1 - \delta^{1} = \hat{\eta}/\beta$ where  $\hat{\eta}$ is $\mathcal{O}(1)$.
To achieve nontrivial results $m$ must also be scaled with $\beta$ and we 
reparametrize $m = \hat{m}/\beta$. Then for $\alpha > \alpha_{\rm c}(\gamma)$ 
the zero temperature free energy functional is given by:
\begin{eqnarray}
\frac{2 F}{N K} &=& \alpha \left\{ 
\frac{1 - 2 z(\Delta) + z(\delta^{0}) + 3 \gamma^{2} \pi/(\pi - 3)
}{\kappa \hat{\eta} + \hat{m}( 1 - z(\delta^{0})) + 3 \pi / (\pi - 3)} \right.
\nonumber \\
& + & \left. \frac{1}{\hat{m}} \ln \left[
\frac{\kappa \hat{\eta} + \hat{m} ( 1 - z(\delta^{0})) + 3 \pi/(\pi - 3)}{
\kappa \hat{\eta} + 3 \pi/(\pi - 3)} \right] \right\} \nonumber \\ 
& - & \frac{\delta^{0} - (\Delta)^{2}}{\hat{\eta} + \hat{m} ( 1 - \delta^{0})}
- \frac{1}{\hat{m}} \ln \left[ \frac{\hat{\eta} + \hat{m}(1 - \delta^{0})}{
\hat{\eta}} \right] \; ,
\label{free0}
\end{eqnarray}  
where $\kappa = (2 \sqrt{3} - 3)/(\pi - 3)$ and 
$z(x) = (-3/\pi) (x - 2 \arcsin(x/2))$. 
The replica symmetric case may be recovered by either taking the limit
$\hat{m} \rightarrow 0$ or the limit  $\delta^{0} \rightarrow 1$.

These equations still admit an at least metastable unspecialized
solution with $\Delta = 0$ for all $\alpha >\alpha_{\rm c}(\gamma) $. 
But now replica symmetry is broken in this solution, and
this also holds in the noiseless case $\gamma=0$. 
Above a second critical $\alpha$ the stable solution is 
specialized ($\Delta > 0$) and 
remarkably even in the noisy case this specialized solution does not exhibit
replica symmetry breaking.

The lifting of RSB with the onset of specialization is illustrated in
Figure 1 for $\gamma=\gamma_0$. 
Fixing $\Delta$ and maximizing (\ref{free0}) w.r.t.
to the remaining order parameters corresponds to calculating the free
energy of a system with a state space constrained to vectors yielding a 
specialized student/teacher overlap of $\Delta$. At the maximum
$F/P$ is the training error of the constrained system shown in Figure 1.

\begin{figure}
\label{bildchen}
\begin{center}
\begin{picture}(10,3.6)(0,0)
\put(4,0){$\Delta$}
\put(9.25,0){$\alpha$}
\put(0.25,3.4){$\epsilon_{t}$}
\put(5.5,3.4){$\epsilon_{t}$}
\put(1,2.5){\tiny{RSB}}
\put(1,1.2){\tiny{RS}}
\put(6.1,2.3){\tiny{$\Delta = 0$, RSB}}
\put(6.1,2.1){\tiny{$\Delta = 0$, RS}}
\put(8.5,1.9){\tiny{$\Delta > 0$}}
\put(1.35,2.5){\vector(1,-1){0.3}}
\put(1.28,1.2){\vector(1,-1){0.25}}
\put(7.1,2.3){\vector(2,-1){0.4}}
\put(6.95,2.1){\vector(2,-1){0.5}}
\put(9,2.05){\vector(1,2){0.25}}
\put(0,-0.6){\resizebox{0,475\textwidth}{!}{\includegraphics{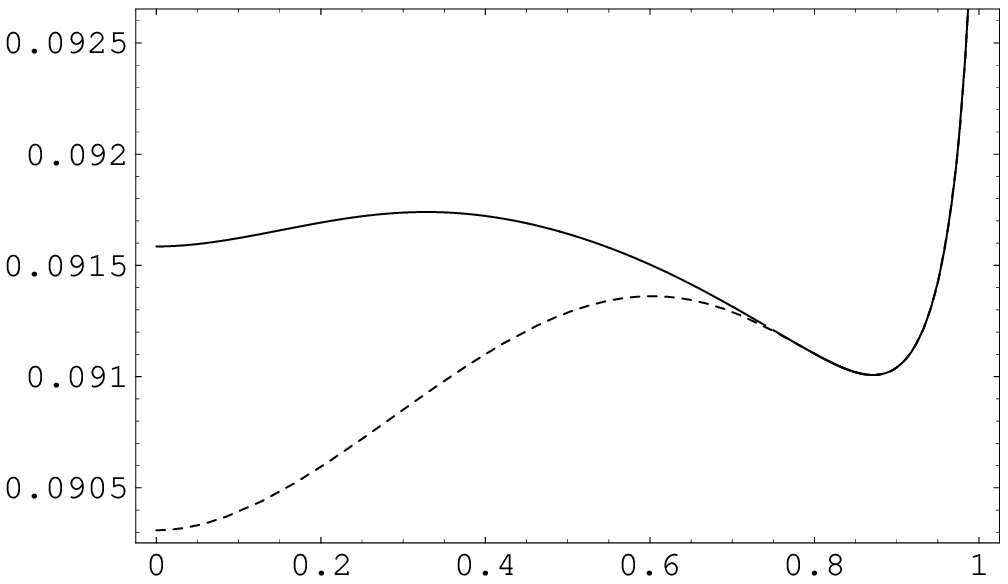}}}
\put(5.25,-0.6){\resizebox{0,475\textwidth}{!}{\includegraphics{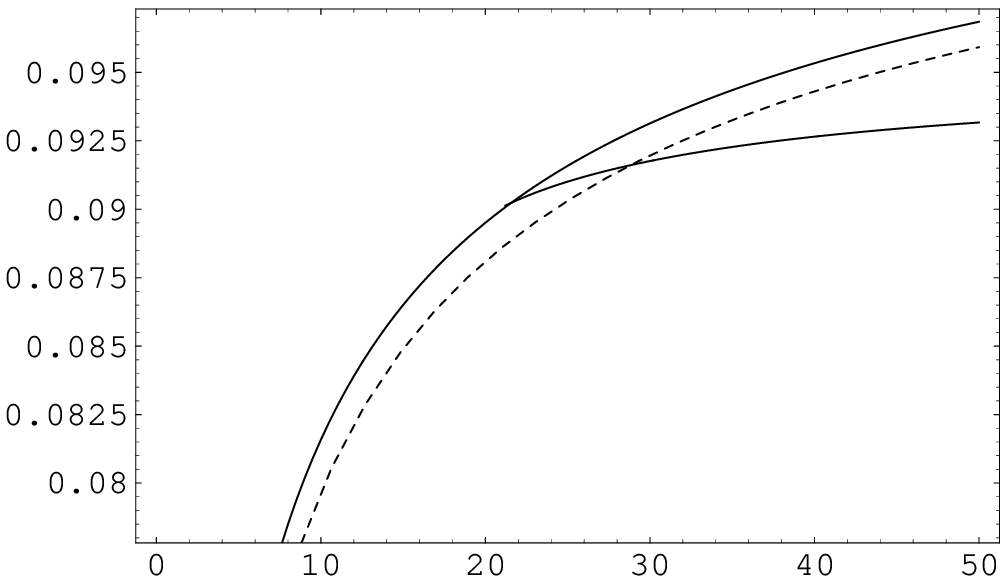}}}
\end{picture}
\end{center}
\caption{Results for the classification problem and for the noisy regression 
problem with $\gamma = \gamma_0$. 
Left panel: $\epsilon_{t}(\Delta)$ for 
$\alpha = 25$. Dashed line: replica symmetrical solution, solid line: 
one-step replica symmetry breaking solution. Right panel: 
$\epsilon_{t}(\alpha)$. At $\alpha_{\rm c}(\gamma_0) \approx 0.3$
replica symmetry is broken. At $\alpha \approx 21.5$ replica symmetry
is restored with the onset of specialization. The dashed line shows the 
(wrong) results of a replica symmetrical calculation. }
\end{figure}

The physically relevant states in the case of training with unconstrained
$\Delta$ are given by the minima of these curves. Both in the RS and the 
RSB parametrizations  we find a local minimum at $\Delta = 0$, 
which corresponds to a metastable  unspecialized configuration of the system. 
Here the RSB solution yields a greater free energy than the RS solution and 
therefore is the only physically relevant solution.

With increasing  $\Delta$ both curves approach each other, and 
the RSB and RS solutions merge at
$\Delta \approx 0.78$, i.e. for sufficiently large $\Delta$ there is 
no replica symmetry breaking.

There is a second minimum of the free energy at $\Delta \approx 0.87$
which corresponds to a replica symmetrical specialized phase of the 
learning with unconstrained $\Delta$ which yields a lower 
free energy than the unspecialized solution and therefore is the 
globally stable configuration.

In general, we find the following scenario which is illustrated in the
right panel of Figure 1 for $\gamma=\gamma_0$.
 For all values of $\alpha$ there
is an unspecialized solution with constant prediction error 
($\epsilon_{p} = 1/3 - 1/\pi + \gamma^{2}/2$). 
Replica symmetry is broken in this solution for 
$\alpha > \alpha_{\rm c}(\gamma)$. Beyond a second critical $\alpha$
the unspecialized solution is only metastable and the stable solution
is specialized and replica symmetric. In the noiseless case, the two
critical values of $\alpha$ coincide, and thus replica symmetry is never
broken in the stable state. In the noisy case, the prediction error
decays as $1/\alpha$ to its asymptotical value $\gamma^{2}/2$
in the specialized phase.
  
For the classification problem, $\gamma = \gamma_0$, the $1/\alpha$
decay of the prediction error translates into the following asymptotics
of the classification error:
\begin{equation}
\epsilon_{c} \sim \frac{\sqrt{\frac{\pi}{2} - 1}}{\pi^{\frac{1}{4}}}
\frac{1}{\sqrt{\alpha}} \; . 
\end{equation}

This slow decay of $\epsilon_{c}$ reflects the cost of treating
the classification problem as a regression problem and thus mapping a 
realizable case onto an unrealizable one. Based on the results of
\cite{Schwartze} one would expect a $1/\alpha$ asymptotics of the
classification error, if the hard cost function (\ref{emp}) would be used
instead of the quadratic deviation (\ref{energie}). 
Thus future research into batch
learning should investigate the use of cost functions like the ones proposed in
\cite{Barkai:Seung:Sompolinsky} for the online scenario.

For the general case of noisy regression, it is remarkable that replica
symmetry breaking is only a transient phenomenon in that the specialized
state which is the stable one for large $\alpha$ is replica symmetric even in
this unrealizable scenario. 

Part of this work was carried out during the Seminar on Statistical 
Physics of Neural Networks
at the Max-Planck-Institut f\"ur Physik komplexer Systeme
in Dresden. Further this work was supported by a British Council grant
(British-German Academic Research Collaboration, programme project 1037)
and a DAAD grant (project number 9818105). 

We thank Georg Reents and Enno Schl\"osser for stimulating
discussions and a critical reading of the manuscript.

\end{document}